\documentclass[epj]{webofc}
\usepackage[varg]{txfonts}   % Web of Conferences font
\usepackage{hyperref}
\wocname{EPJ Web of Conferences}
\woctitle{CONF12}

\newcommand{\beq}{\begin{equation}}
\newcommand{\eeq}{\end{equation}}
\newcommand{\bea}{\begin{array}}
\newcommand{\eea}{\end{array}}
\newcommand{\beqa}{\begin{eqnarray}}
\newcommand{\eeqa}{\end{eqnarray}}

\def\beqa{\begin{eqnarray}}
\def\eeqa{\end{eqnarray}}

\begin{document}
\selectlanguage{english}
\title{
Study of lattice QCD at finite baryon density using the canonical approach
}

\author{V.~G.~Bornyakov\inst{1,2,3} \and
        D.~L.~Boyda\inst{4,2} \and
        V.~A.~Goy\inst{4,2} \and
        A.~V.~Molochkov\inst{3,2} \and
        Atsushi Nakamura\inst{3,5,6} \and
        A.~A.~Nikolaev\inst{3,2}\fnsep\thanks{\email{nikolaev.aa@dvfu.ru}} \and
        V.~I.~Zakharov\inst{2,3,7}
}

\institute{Institute for High Energy Physics NRC ``Kurchatov Institute'', 142281 Protvino, Russia 
\and
	Institute for Theoretical and Experimental Physics NRC ``Kurchatov Institute'', 117218 Moscow, Russia
\and
	School of Biomedicine, Far Eastern Federal University, 690950 Vladivostok, Russia
\and
	School of Natural Sciences, Far Eastern Federal University, 690950 Vladivostok, Russia
\and
	Research Center for Nuclear Physics (RCNP), Osaka University, Ibaraki, Osaka, 567-0047, Japan
\and
	Theoretical Research Division, Nishina Center, RIKEN, Wako 351-0198, Japan
\and
	Moscow Institute of Physics and Technology, 141700 Dolgoprudny, Moscow Region, Russia
}

\abstract{%
At finite baryon density lattice QCD first-principle calculations can not be
performed due to the sign problem. In order to circumvent this problem, we
use the canonical approach, which provides reliable analytical continuation
from the imaginary chemical potential region to the real chemical potential region.
We briefly present the canonical partition function method, describe our formulation, and
show the results, obtained for two temperatures: $T / T_c = 0.93$ and
$T / T_c = 0.99$ in lattice QCD with two flavors of improved Wilson fermions.
}
\maketitle
\section{Introduction}
%---------------------
\label{sec:introduction}

The  phase structure of hadronic matter at finite temperature and density
is studied in experiments at most important modern accelerators  RHIC (BNL)~\cite{Adams:2005dq}, LHC (CERN)~\cite{Aamodt:2008zz}  and future experiments
 FAIR (GSI) and NICA (JINR)  will be devoted to such studies.
The lattice QCD numerical simulations have a mission to provide theoretical results from the first principle
calculations.
Indeed, at finite temperature with zero chemical potential,
the phase structure was satisfactorily investigated.
But it is very difficult to study the finite density regions by the lattice QCD
because of the sign problem:
the fermion determinant $\det\Delta(\mu_B)$ at non-zero baryon chemical potential $\mu_B$ is in general not real.
It is impossible to apply standard Monte-Carlo techniques to computations with the partition function
\beq
Z_{GC}(\mu_q,T,V) = \int \mathcal{D}U (\det\Delta(\mu_q))^{N_f} e^{-S_G},
\label{Eq:PathIntegral}
\eeq
where $S_G$ is a gauge field action, $\mu_q=\mu_B/3$ is quark chemical potential,   $T=1/(aN_t)$ defines temperature, $V=(aN_s)^3$ is volume, $a$ is lattice spacing, and $N_t,\,N_s$ - number of lattice sites in time and space directions respectively.
There have been many trials, see reviews~\cite{Muroya:2003qs,Philipsen:2005mj,deForcrand:2010ys}, and yet it is still very hard to get reliable results at $\mu_B/T>1$.
In this study we apply canonical approach. The canonical approach was studied in a number of papers~\cite{deForcrand:2006ec,Ejiri:2008xt,Li:2010qf,Li:2011ee,Danzer:2012vw,Gattringer:2014hra,
Fukuda:2015mva,Nakamura:2015jra}. We suggest new method to compute canonical partition function $Z_C(n, T, V)$, which allows to compute it for large values of $n$, where $n$ is a net number of quarks and antiquarks. Our results
for $Z_C(n, T, V)$ obtained with the new method are in a good agreement with results obtained with the known method of hopping parameter expansion.

\section{Details of the new method}
\label{sec:approach}

The canonical approach is based on the following relations. First, this is a relation between the grand canonical partition function $Z_{GC}(\mu_q, T, V)$ and the canonical one $Z_C(n, T, V)$:
\begin{eqnarray}
Z_{GC}(\mu, T, V)=\sum_{n=-\infty}^\infty Z_C(n,T,V)\xi^n,
\quad
\label{ZG}	
\end{eqnarray}
where $\xi=e^{\mu_q/T}$ is the fugacity and the eq.~(\ref{ZG})
is called fugacity expansion.
The inverse of this equation can be presented in the following form~\cite{Hasenfratz:1991ax}:
\begin{eqnarray}
Z_C\left(n,T,V\right)=\int_0^{2\pi}\frac{d\theta}{2\pi}
e^{-in\theta}Z_{GC}(\theta,T,V)\,.
\label{Fourier}
\end{eqnarray}
In the right hand side of eq.~(\ref{Fourier}) we  see the grand canonical partition function  $Z_{GC}(\theta,T,V)$ for imaginary chemical potential $\mu_q=i\mu_{qI} \equiv iT\theta$. It is known that standard Monte-Carlo simulations are possible for this partition function since the fermionic determinant is real for imaginary $\mu_q$.

The QCD partition function $Z_{GC}$ for imaginary chemical potential is a periodic function of $\theta$: $Z_{GC}(\theta) = Z_{GC}(\theta+2\pi/3)$. This symmetry is called Roberge-Weiss symmetry~\cite{Roberge:1986mm}. As a consequence of this periodicity the canonical partition functions $Z_C(n,T,V)$  are nonzero only for $n=3k$. QCD possesses a rich phase structure at non-zero $\theta$, which depends on the number of flavors $N_f$ and the quark mass $m$~\cite{Bonati:2014kpa}.

Quark number density $n_q$ for $N_f$ degenerate quark flavours is defined by the following equation:
\beq
\frac{n_{q}}{T^{3}} = \frac{1}{VT^{2}}\frac{\partial}{\partial \mu_q}\ln
Z_{GC} = \frac{N_{f}N_{t}^{3}}{N_s^3 Z_{GC}} \int \mathcal{D}U e^{-S_G} (\det\Delta(\mu_q))^{N_f}
\mathrm{tr}\left[\Delta^{-1}\frac{\partial \Delta}{\partial (\mu_q / T)}\right]\,.
\label{density1}
\eeq
It can be computed numerically for imaginary chemical potential. Note, that for the imaginary chemical potential $n_q$ is also purely imaginary: $n_q = i n_{qI}$.

From eqs.~(\ref{ZG}) and~(\ref{density1}) it follows that densities $n_{q}$ and $n_{qI}$ are related  to $Z_C(n,T,V)$ (below we will use the notation $Z_n$ for the ratio
$Z_C(n,T,V) / Z_C(0,T,V)$) by equations
\beq
\label{density_re}
n_{q}/T^3  = {\cal{N}}\frac{2\sum_{n>0} n Z_n \sinh(n\theta)}{1+2\sum_{n>0} Z_n \cosh(n\theta)}\,,
\eeq
\beq
\label{density2}
n_{qI}/T^3  = {\cal{N}}\frac{2\sum_{n>0} n Z_n \sin(n\theta)}{1+2\sum_{n>0} Z_n \cos(n\theta)}\,,
\eeq
where ${\cal{N}} = N_t^3 / N_s^3$ is a normalization constant. With the help of equations~(\ref{density_re}) and~(\ref{density2}) one can construct quark number density at real and imaginary chemical potential once $Z_n$ are known.

One can compute $Z_{GC}(\theta,T,V)$ using numerical data for $n_{qI}/T^3$ via numerical integration over imaginary chemical potential:
\beq
L_Z(\theta) \equiv \log\frac{Z_{GC}(\theta,T,V)}{Z_{GC}(0,T,V)}  = - V \int_{0}^{\theta} d \tilde{\theta}~n_{qI}(\tilde{\theta})\,,
\label{integration_1}
\eeq
where we omitted $T$ and $V$ from the grand canonical partition function notation. Then $Z_n$ can be computed as
\beq	
Z_n = \frac{\int_0^{2\pi}\frac{d\theta}{2\pi} e^{-in\theta} e^{L_Z(\theta)} }{ \int_0^{2\pi}\frac{d\theta}{2\pi}
 e^{L_Z(\theta)} }
\label{Fourier_2}
\eeq

In the present work we use modified version of this approach.
Instead of numerical integration in~(\ref{integration_1})
we fitted $n_{qI}/T^3$ to theoretically motivated functions
of $\mu_{qI}$. It is well known that in the confining phase the hadron resonance gas model provides good description of the chemical potential dependence of thermodynamic observables~\cite{Karsch:2003zq}. Thus it is reasonable to fit the density to a Fourier expansion
\beq
n_{qI}(\theta)/T^3 = \sum_{n=1}^{n_{max}} f_{3n} \sin(3n \theta)\,.
\label{eq_fit_fourier}
\eeq
This type of the fit was used in Ref.~\cite{Takahashi:2014rta} and conclusion was made that it works well. As an alternative one can also try to fit $n_{qI}$ to an odd power polynomial of $\theta$:
\beq
n_{qI}(\theta)/T^3 = \sum_{n=1}^{n_{max}} a_{2n-1} \theta^{2n-1}\,.
\label{eq_fit_polyn}
\eeq

\section{Numerical results}
\label{simulation}
To demonstrate our method we make simulations of the lattice QCD with $N_f = 2$ clover improved Wilson quarks and Iwasaki improved gauge field action:
\begin{eqnarray}
  S   &=& S_G + S_F, \\
  S_G &=&
  -{\beta}\sum_{x,\mu\nu}  \left(
   c_0 W_{\mu\nu}^{1\times1}(x)
   + c_1 W_{\mu,\nu}^{1\times2}(x) \right), \\
  S_F &=& \sum_{f=u,d}\sum_{x,y}\bar{\psi}_x^f \Delta_{x,y}\psi_y^f,
  \label{eq:action}
\end{eqnarray}
with $\beta=6/g^2$, $c_1=-0.331$, $c_0=1-8c_1$, $W_{\mu\nu}$ denoting the plaquettes, and
\begin{eqnarray}
 \Delta_{x,y} = \delta_{xy} &-&\kappa\sum_{i=1}^3 \{(1-\gamma_{i})U_{x,i}\delta_{x+\hat{i},y} + (1+\gamma_{i})U_{y,i}^{\dagger}\delta_{x,y+\hat{i}}\}
       \nonumber \\ 
   &-&{\kappa} \{e^{a\mu_q}(1-\gamma_{4})U_{x,4}\delta_{x+\hat{4},y} + e^{-a\mu_q}(1+\gamma_{4})U_{y,4}^{\dagger}\delta_{x,y+\hat{4}}\} - \delta_{xy}{c_{SW}}{\kappa}\sum_{\mu<\nu}\sigma_{\mu\nu}P_{\mu\nu}\,,
\label{eq:fermact}
\end{eqnarray}
where $P_{\mu\nu}$ is the clover definition of the lattice field strength tensor and $c_{SW} =(W^{1\times 1})^{-3/4}=(1-0.8412\beta^{-1})^{-3/4}$ is the Sheikholeslami–Wohlert coefficient.

We simulate $16^3 \times 4$ lattices at temperatures $T/T_c=0.99$ and 0.93 in the confinement phase along the line of constant physics with $m_{\pi}/m_{\rho}=0.8$. All parameters of the action, including $c_{SW}$ value, were borrowed from the WHOT-QCD collaboration paper~\cite{Ejiri:2009hq}. We compute the number density on samples of $N_{conf}$ configurations with $N_{conf}=1800$, using every 10-th trajectory produced with Hybrid Monte-Carlo algorithm.
\begin{figure}[htb]
\begin{center}
	\begin{minipage}[t]{0.49\textwidth}
    	\includegraphics[width=0.95\textwidth,angle=0]{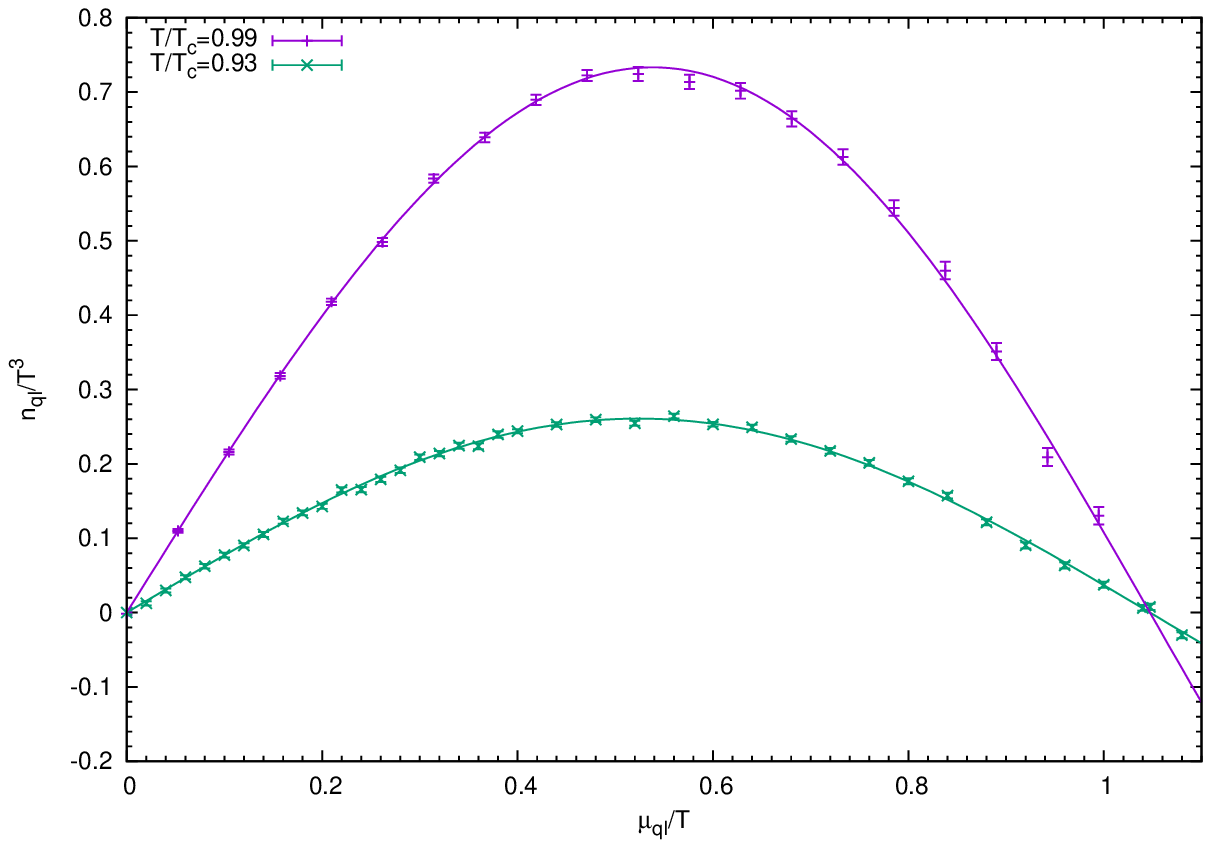}
\caption{Imaginary density as a function of $\theta$ for two temperatures in the confining phase. The curves show fits to the function~(\ref{eq_fit_fourier}) with $n_{max} = 1$ for $T/T_c = 0.93$ and $ n_{max}=2$ for $T/T_c = 0.99$.}
\label{density_conf}
    \end{minipage}
\hfill
    \begin{minipage}[t]{0.49\textwidth}
    \vspace{-50mm}
	\includegraphics[width=0.75\textwidth,angle=-90]{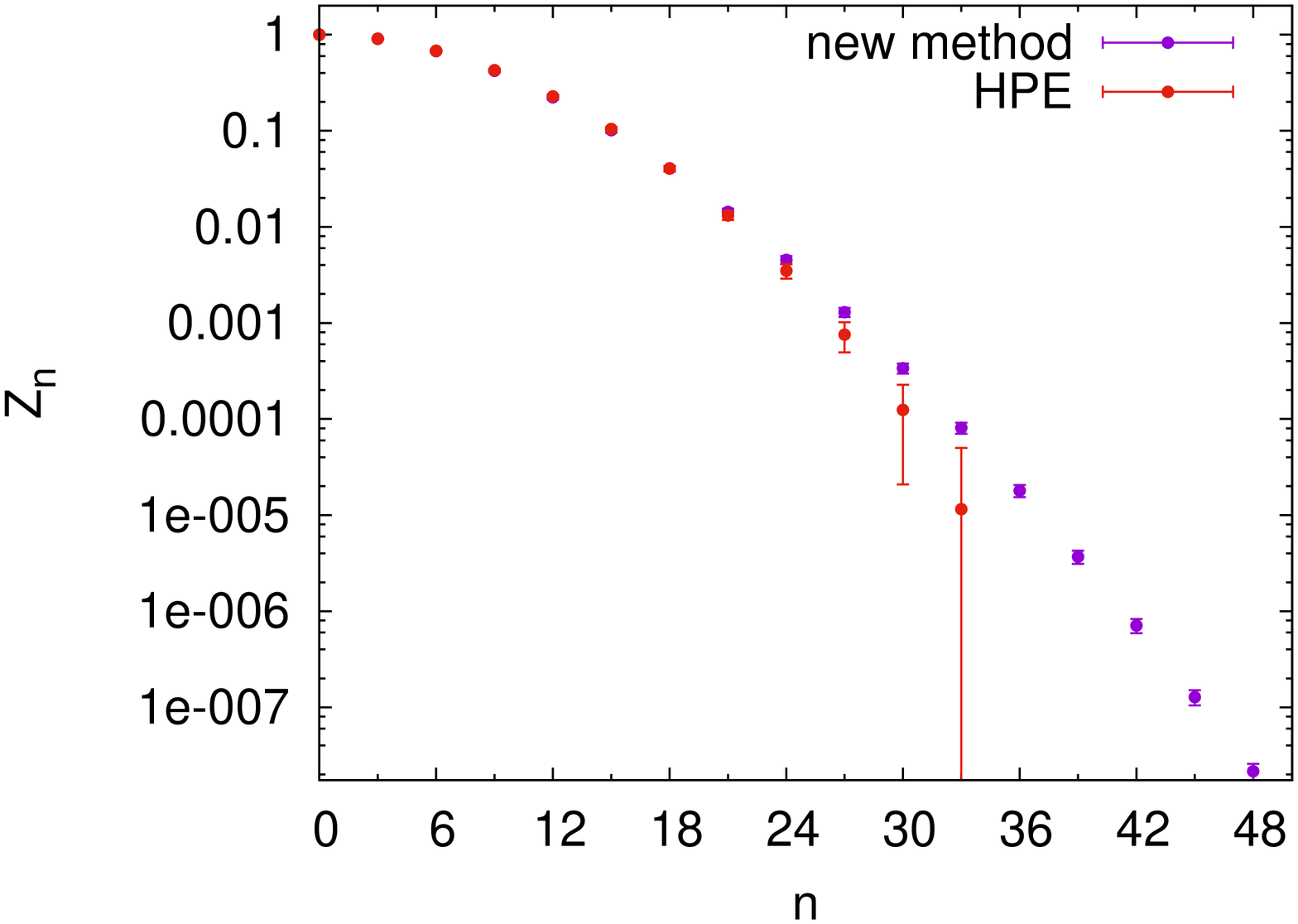}
	\caption{Dependence of $Z_n$ on $n$ for $T/T_c=0.93$ for two methods of computation.}
	\label{Zn_conf_compar}
    \end{minipage}
\end{center}
\end{figure}

We employ the hopping parameter expansion to compute $Z_n$ and compare with $Z_n$ values obtained with our new method. The Wilson Dirac operator from~(\ref{Eq:PathIntegral}) may be written in the form
\beq
\label{eq:D_op_Q}
\Delta = I - \kappa Q\,,
\eeq
both in the case of standard and clover improved Wilson fermions. Then one can rewrite the fermionic determinant in the following way~\cite{Nakamura:2015jra}:
\beq
\label{eq:Det_expansion}
\det\Delta = exp\Bigl[ \mathrm{Tr}\,ln \left( I - \kappa Q \right) \Bigr] = exp\Bigl[ - \sum_{n = 1}^{\infty} \frac{\kappa^n}{n} \mathrm{Tr} Q^n \Bigr]\,.
\eeq
The expansion in~(\ref{eq:Det_expansion}) is in fact expansion over the closed paths on the lattice, and thus it can be rewritten as
\beq
\label{eq:Det_expansion_W_n}
\det\Delta = exp\Bigl[ \sum_{n = - \infty}^{\infty} W_n \xi^n \Bigr]\,,
\eeq
where $n$ is number of windings in the temporal direction, $W_n$ are complex coefficients which are called winding numbers. They satisfy the property $W_{-n} = W_n^*$.
In the case of the imaginary chemical potential the formula~(\ref{eq:Det_expansion_W_n}) will become
\beq
\label{eq:Det_expansion_im_muq}
\det\Delta = e^{W_0} e^{2\sum_{n = 1}^{\infty} \Bigl( \text{Re}[W_n]\cos (n \theta) - \text{Im}[W_n]\sin (n \theta) \Bigr) }\,.
\eeq
It is important to note that hopping parameter expansion~(\ref{eq:Det_expansion}) converges properly only for heavy quark masses~\cite{Nakamura:2015jra}. Our simulations were performed for the quark masses from this range.

Below we present our results obtained below $T_c$. In Fig.~\ref{density_conf} we show $n_{qI}$ for $\theta \in [0;\pi/3]$ together with fits to eq.~(\ref{eq_fit_fourier}), the fit results are presented in Table~\ref{table_results_2}. We found good fit with $n_{max}=1$ for $T/T_c = 0.93$ while for $T/T_c=0.99$ fit with $n_{max}=2$ is necessary. In Table~\ref{table_results_2} we also show coefficients $a_1$ and $a_3$ of the Taylor expansion of eq.~(\ref{eq_fit_polyn}) as well as the respectve results from~\cite{Ejiri:2009hq} (two last columns). We observe good agreement for the first Taylor expansion coefficient within error bars and substantially smaller error bars in our study than in~\cite{Ejiri:2009hq}. For the second coefficient we see a noticeable disagreement, which means that at $T < T_c$ the fit~(\ref{eq_fit_polyn}) works well only up to the first order and at small enough chemical potential values.

Eq.~(\ref{Fourier_2}) to compute $Z_n$ now looks as follows:
\beq
Z_{3n} = \frac{ \int_0^{2\pi}\frac{d\theta}{2\pi} \cos(3n\theta) e^{\sum_{m=1}^{n_{max}} \tilde{f}_{3m} \cos(3m \theta) }}
{\int_0^{2\pi}\frac{d\theta}{2\pi} e^{ \sum_{m=1}^{n_{max}}  \tilde{f}_{3m} \cos(3m \theta)}} = \frac{ \int_0^{6\pi}\frac{dx}{6\pi} \cos(nx) e^{\sum_{m=1}^{n_{max}} \tilde{f}_{3m}  \cos(mx) }}
{\int_0^{6\pi}\frac{dx}{6\pi}  e^{\sum_{m=1}^{n_{max}} \tilde{f}_{3m} \cos(mx)}}\,,
\label{Fourier_conf}
\eeq
where $\tilde{f}_{3n} = (N_s^3 f_3) / (N_t^3 3n)$. In the case $n_{max} = 1$ this can be expressed as
\beq
Z_{3n} = I_n(\tilde{f}_3) / I_0(\tilde{f}_3)\,,
\label{Z_3n_Bessel}
\eeq
where $I_n(x)$ is the modified Bessel function of the first kind. Results for $Z_n$ at $T/T_c = 0.93$ are presented in Fig.~\ref{Zn_conf_compar}.

We first check if computed $Z_n$ reproduce the data for $n_{qI}/T^3$ and find nice agreement between data and values of $n_{qI}/T^3$ computed via eq.~(\ref{density2}). The deviation for the full interval [0.0; $\pi/3$] is less than $0.3\%$. Next we compare with hopping parameter expansion, respective results are also presented in Fig.~\ref{Zn_conf_compar}. We used full statistics (1800 configurations at $\mu_{qI} = 0$) and $W_n$ up to $n=15$ in eq.~(\ref{eq:Det_expansion_W_n}) for this computation.
One can see agreement between two results up to $n=21$.
We believe that the disagreement for higher $n$ is explained by inaccuracy in computation of $Z_n$ computed by HPE.
The statistical errors for $Z_n$ grow very fast with $n$. It is necessary to improve the HPE method accuracy before the
conclusion about agreement at large $n$ can be made.
\begin{table*}[th]
\begin{center}
\begin{small}
\begin{tabular}{|c|c|c|c|c|c|c|c|} \hline
$T/T_c$  & $f_3$ & $ f_6$& $ a_1$ &$a_3$& $\chi^2/N_{dof}, N_{dof}$  & $ 2c_2$ & $ 24 c_4$  \\ \hline
0.99    &  0.7326(25)&-0.0159(21)& 2.102(5)    &-2.719(17)& 0.83, 18 &  2.071(34) & 17.4(47) \\
0.93    & 0.2608(8) & -          & 0.7824(24) &-1.1736(36)& 0.93, 37 &  0.713(40) & 2.0(48) \\ \hline
\end{tabular}
\end{small}
\end{center}
\caption{ Results of fitting data for $n_{qI}/T^3$ in the confinement phase to the functions~(\ref{eq_fit_fourier}) and~(\ref{eq_fit_polyn}). The 6th column with $\chi^2$ and $N_{dof}$ demonstrates the results for the fit~(\ref{eq_fit_fourier}).}
\label{table_results_2}
\end{table*}

\begin{figure}[htb]
\begin{center}
    \begin{minipage}[t]{0.49\textwidth}
	\includegraphics[width=0.95\textwidth]{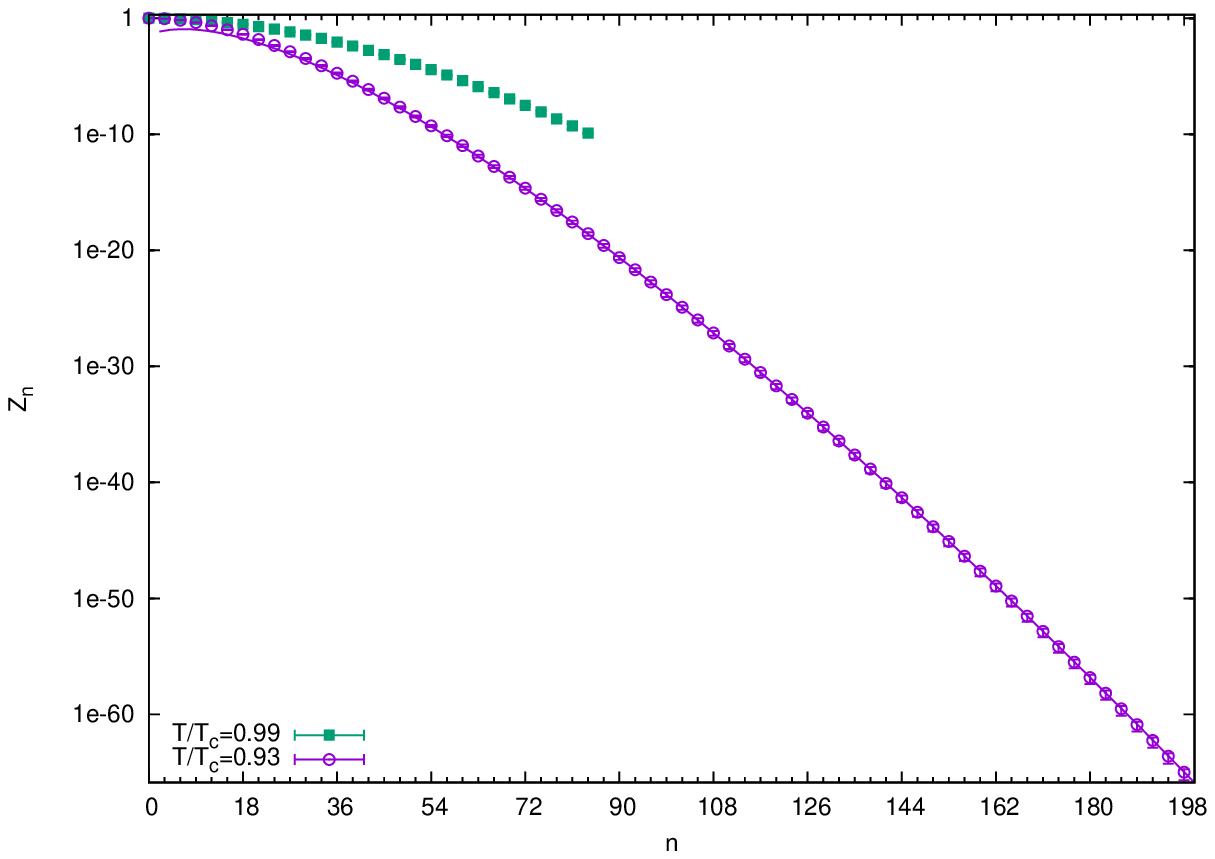}
	\caption{$Z_n$ vs $n$ for $T/T_c=0.93$ and 0.99. The curve shows results of fitting to asymptotic behaviour~(\ref{asymptotics-2}) for $T/T_c=0.93$.}
	\label{Zn_conf_all}
    \end{minipage}
\hfill
	\begin{minipage}[t]{0.49\textwidth}
    	\includegraphics[width=0.95\textwidth]{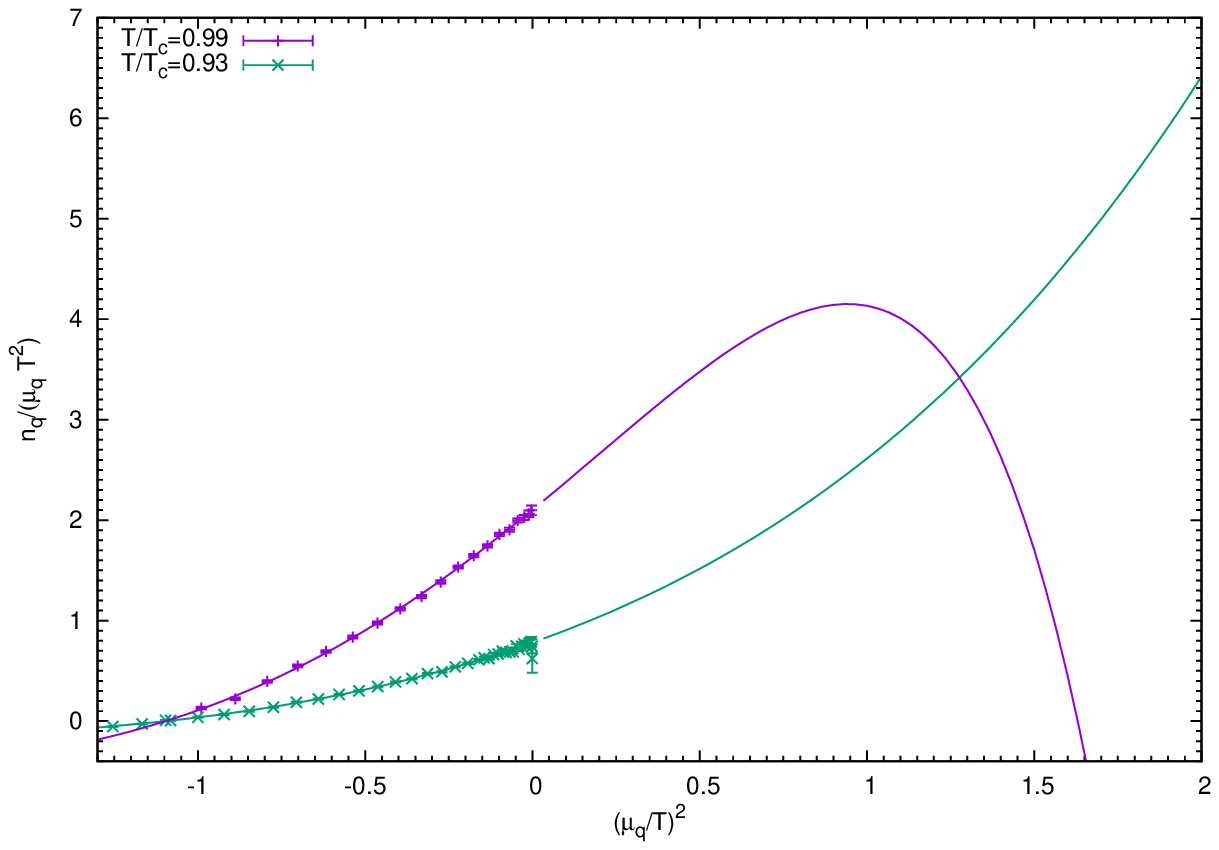}
\caption{Analytical continuation for the number density vs. $\mu_q^2$ for all temperature values.
The curves show respective fits, details of the fits are declared in Table~\ref{table_results_2}.}
\label{dens_anal_cont}
    \end{minipage}
\end{center}
\end{figure}

Still our result indicates that at $T/T_c=0.93$ the fit function~(\ref{eq_fit_fourier}) provides correct values of
$Z_n$ and thus its analytical continuation should be valid up to  values of $\mu_q/T$ beyond Taylor expansion validity
range. Precise determination
of the range of validity of this analytical continuation will be made in future after getting more precise results for the HPE method.

Let us note that one can derive a recursion relation for $Z_n$ when $n_{qI}$ is presented by the function~(\ref{eq_fit_fourier}) with finite $n_{max}$ (for the derivation see Appendix). In particular for $n_{max}=1$ the recursion is just a recursion for $I_n$ which is of the form
\beq
  \tilde{f_3}( Z_{3(n-1)} - Z_{3(n+1)} ) =  2n Z_{3n}.
\label{recursion_1}
\eeq

Also one can get asymptotics for $Z_n$ at large $n$. For $n_{max}=1$ it is
\beq
Z_{3n} = B \frac{(\tilde{f}_3/2)^n}{n !},
\label{asymptotics-2}
\eeq
where $B$ is some  constant. This asymptotics is shown in Fig.~\ref{Zn_conf_all} for
$T/T_c=0.93$ with constant $B$ obtained by the fitting over the range $400 < n < 600$:
$B=0.02813((1)$ for $T/T_c=0.93$. For $n_{max}=N$ it is different:
\beq
Z_{3n} = B\frac{(\tilde{f}_{3N})^{n/N}}{\Gamma(n/N+1)},
\label{asymptotics-3}
\eeq
see Appendix for the derivation. From this asymptotics it follows in particular that the coefficient
$f_{3N}$ has to be positive, otherwise the condition of positivity of $Z_{3n}$ will not hold.
Our current fitting function for $T/T_c=0.99$ which has $f_{3N} \equiv f_6<0$ does not satisfy
this requirement. We need to improve statistics for this temperature to obtain the coefficient
for the next harmonics in eq.~(\ref{eq_fit_fourier}). Evidently with present fitting function we can not go to large values of $\mu_q$ for this temperature.

At the end of this section we show in Fig.~\ref{dens_anal_cont} the ratio $n_q / (\mu_q T^2)$ as a function of $\mu_q^2$ for negative and positive values. This way of presentation, borrowed from~\cite{Gunther:2016vcp}, allows to show in one plot the simulation results obtained at  $\mu_q^2 < 0$ and analytical continuation of our fitting functions to $\mu_q^2 > 0$. For the temperature $T/T_c = 0.93$ results seem to be reasonable, but for the temperature $T/T_c=0.99$ we need to improve the statistics.

\section{Conclusions}
\label{conclusions}

We have presented new method to compute the canonical partition functions $Z_n$.
It is based on the fitting of the imaginary number density for all values of imaginary
chemical potential to the theoretically motivated fitting function, which is Fourier-type fit~(\ref{eq_fit_fourier}) in the case of confinement phase.

Using fit results we compute canonical partition functions $Z_n \equiv Z_C(n,T,V) / Z_C(0,T,V)$ at $T/T_c = 0.93$ and 0.99 via Fourier transformation~(\ref{Fourier_2}). It was necessary to use the multi-precision library~\cite{FMlib} to compute $Z_n$ which change over many orders of magnitude. For both temperatures we have checked that precision of computation of $Z_n$ was high enough to reproduce the imaginary number density $n_{qI}$ via eq.~(\ref{density2}).

At the temperature $T/T_c = 0.93$ we compared our results for $Z_n$ with $Z_n$ computed by the hopping parameter expansion. We found that two sets of $Z_n$ computed by completely independent methods agree well, see the Fig.~\ref{Zn_conf_compar}. This means that the fitting function used is a proper approximations for the imaginary number density in the full range of $\mu_{qI}$ values. Furthermore, this means that the analytical continuation to the real chemical potential can in principle be done beyond the Taylor expansion validity range since this analytical continuation of the quark number density coincides with $n_q$ computed with the help of correctly determined $Z_n$ via eq.~(\ref{density_re}). Thus the new method in principle allows to compute the number density $n_q$ beyond Taylor expansion.

Note that our new method is not limited to the heavy quark mass values like HPE, nor small $\mu$ values like Taylor expansion. Once we calculate $Z_n$ using new method, we can calculate many thermodynamical
quantities, {\it i.e.} pressure, number density and its higher moments. 

Using our results for the number density $n_{qI}$ we computed the Taylor expansion coefficients for the number density from which respective coefficients for the pressure may easily be restored. We found good agreement for the first coefficient with earlier results obtained in~\cite{Ejiri:2009hq} via direct computation of these coefficients. Moreover we found that our error bars for these coefficients are in general substantially smaller than  error bars quoted in~\cite{Ejiri:2009hq}. Thus we confirmed analogous observation made in~\cite{Gunther:2016vcp}.

We obtained asymptotics of $Z_n$ at large $n$ which is defined by the eq.~(\ref{asymptotics-3}). Such a decreasing of $Z_n$ is fast enough to provide convergence of the infinite sums in eqs.~(\ref{ZG}) and~(\ref{density_re}).

\begin{acknowledgement}
This work was completed due to support by
RSF grant under contract 15-12-20008. Computer simulations were performed on the FEFU GPU cluster ``Vostok-1'' and the MSU supercomputer ``Lomonosov''.
\end{acknowledgement}

\appendix
\section*{Appendix: recursion relations for $Z_{3n}$}
\label{appendix}
In this Appendix we derive a recursion relation for $Z_{3n}$ in the confinement phase.
Let us introduce notations $3\theta=x$ and $\tilde{f}_n=f_n/(2C)$.

For the case of $n_{max}>1$ in eq.~(\ref{eq_fit_fourier}) it is possible to derive a recursion relations similar to relation~(\ref{recursion_1}). Let us show this for $n_{max}=2$. We have
\beq
\tilde{f}_3 \sin (x) +  \tilde{f}_6 \sin (2x) = \frac{\sum_n n Z_{3n} \sin (n x)} {1 + 2  \sum_n Z_{3n} \cos( n x)}\,.
\eeq
Then
\beq
( \tilde{f}_3 \sin (x) +  \tilde{f}_6 \sin (2x) ) ( 1 + 2  \sum_n Z_{3n} \cos( n x) ) = \sum_n n Z_{3n} \sin (n x)\,. \nonumber
\eeq

Computing Fourier modes on both sides we get:
\beqa
&&\tilde{f}_3 \int_{-\pi}^{\pi} dx \sin (x) \left( 1 + 2  \sum_n  Z_{3n} \cos( n x) \right) \sin(m x) \nonumber \\
&+& \tilde{f}_6 \int_{-\pi}^{\pi} dx \sin (2x) \left( 1 + 2  \sum_n  Z_{3n} \cos( n x) \right) \sin(m x) =  \int_{-\pi}^{\pi}  dx  \sum_n n Z_{3n} \sin (n x) \sin(m x)
\eeqa

\beq
 \tilde{f}_3 ( Z_{3(m-1)} - Z_{3(m+1)} ) + \tilde{f}_6 ( Z_{3(m-2)} - Z_{3(m+2)} )  =   m Z_{3m}
\eeq
or
\beq
Z_{3(m+2)} =  Z_{3(m-2)} - \frac{m}{\tilde{f}_6} Z_{3m} + \frac{\tilde{f}_3}{\tilde{f}_6} ( Z_{3(m-1)} - Z_{3(m+1)} )\,.
\eeq
We need $f_3$, $f_6$, $Z_3$, $Z_6$ to compute all $Z_{3m},\,m > 2$. The asymptotical behavior in this case is the following:
\beq
Z_{3n} = B\frac{(\tilde{f}_6)^{n/2}}{\Gamma(n/2+1)}\,.
\label{eq:asympt2}
\eeq

It is easy to get the recursion relation and asymptotics for $n_{max}=N$. The recursion
relation in this case is
\beq
Z_{3(n+N)} = Z_{3(n-N)} - \frac{n}{\tilde{f}_{3N}} Z_{3n} + \sum_{m=1}^{N-1}\frac{\tilde{f}_{3(n-m)}}{\tilde{f}_{3N}} ( Z_{3(n-m)} - Z_{3(n+m)} )
\eeq
and the asymptotics is
\beq
Z_{3n} = B\frac{(\tilde{f}_{3N})^{n/N}}{\Gamma(n/N+1)}\,.
\label{eq:asymp_N}
\eeq
Some conclusions might be drawn from the expression~(\ref{eq:asymp_N}). Firstly, the asymptotics is determined by the highest mode thus $f_{3N}$ has to be positive. Secondly, decreasing of $Z_n$ becomes weaker with increasing of $N$.

The numerical data for $n_{qI}(\theta)$ indicate that in the confinement phase the number of modes $N$ necessary
to describe the data is finite. Then the above considerations apply and we can make a statement that
the radius of convergence is infinite. In the deconfinement phase at temperatures
$T>T_{RW}$ where the first order Roberge-Weiss transition takes place $N$ is definitely infinite. In the range of temperature $T_c < T < T_{RW}$ the situation is unclear.
% BibTeX or Biber users please use (the style is already called in the class, ensure that the "woc.bst" style is in your local directory)
\bibliography{citations_fdensity}
\end{document}